\documentclass[a4paper, 11pt, oneside]{article}
\usepackage[utf8]{inputenc}
\usepackage[english]{babel}
\usepackage{hyperref}
\usepackage{geometry}
\usepackage{amsmath}
\usepackage{amssymb}
\usepackage{amsthm}
\usepackage{cite}
\usepackage{mathrsfs}
\usepackage{graphicx}
\usepackage{wrapfig}
\usepackage{subfig}
\usepackage{textcomp}
\usepackage{xcolor} 
\usepackage{booktabs}
\usepackage{caption}
\usepackage[output-decimal-marker={,}]{siunitx}
\usepackage{listings}
\usepackage[affil-it]{authblk}

\newcommand{\per}{\cdot}
\newcommand{\link}[1]{\footnote{\href{#1}{#1}}}

\definecolor{codescomments}{rgb}{0,0.6,0}
\definecolor{codegray}{rgb}{0.5,0.5,0.5}
\definecolor{codesstrings}{rgb}{1,0.6,0.0}
\definecolor{backcolour}{rgb}{0.82,0.8,0.75}
\definecolor{codeskeywords}{rgb}{0.25,0.5,1}

\lstdefinestyle{mystyle}{
    backgroundcolor=\color{backcolour},   
    commentstyle=\color{codescomments},
    keywordstyle=\color{codeskeywords},
    numberstyle=\tiny\color{codegray},
    stringstyle=\color{codesstrings},
    basicstyle=\ttfamily\footnotesize,
    breakatwhitespace=false,         
    breaklines=true,                 
    captionpos=b,                    
    keepspaces=true,                 
    numbers=left,                    
    numbersep=5pt,                  
    showspaces=false,                
    showstringspaces=false,
    showtabs=false,                  
    tabsize=2,
    upquote=true,
    frame=lines
}

\title{\textbf{Deep Neural Network as an alternative to Boosted Decision Trees for PID}}
\date{}
\author{Denis Stanev \footnote{denis.y.stanev@gmail.com} , Riccardo Riva \footnote{riccardo.riva.997@gmail.com} , Michele Umassi \footnote{micheleumassi@gmail.com}}
\affil{Dipartimento di Fisica, Sapienza Universita di Roma, Piazzale Aldo Moro 5, I-00185 Roma, Italy}

\begin{document}
\newgeometry{top=2.5cm,bottom=2.3cm,left=2.3cm,right=2.3cm}
\maketitle

\begin{abstract}
In this paper we recreate, and improve, the binary classification method for particles proposed in Roe et al. (2005) paper \cite{2005NIMPA.543..577R}. Such particles are $tau$ neutrinos, which we will refer to as background, and $electronic$ neutrinos: the signal we are interested in. In the original paper the preferred algorithm is a Boosted decision tree. This is due to its low effort tuning and good overall performance at the time. Our choice for implementation is a deep neural network, faster and more promising in performance. We will show how, using modern techniques, we are able to improve on the original result, both in accuracy and in training time.
\end{abstract}

\tableofcontents

\section{Introduction}
In the last decades, as experiments have become more and more complex and the amount of data produced by the detectors continues to increase, machine learning has proven very useful in tackling the large amount of data generated in High Energy Physics experiments \cite{Iiyama_2021} \cite{Kieseler_2020}. \\
In this paper we recreate, and improve, the binary classification method for particles proposed in Roe et al.  (2005) paper \cite{2005NIMPA.543..577R}, to prove how modern machine learning techniques allow for faster and more accurate results. \\
The particles we will be dealing with are $tau$ neutrinos, which we will refer to as background, and $electronic$ neutrinos: the signal we are interested in. In the original paper the preferred algorithm is a Boosted decision tree. This is due to its low effort tuning and good overall performance at the time. Our choice for implementation is a deep neural network, faster and more promising in performance.

\section{Data Analysis} \label{DataAnalysis}
All of the code written for this paper is available in \href{https://github.com/Denis-Stanev/DNN-vs-BDT}{this Github repository ( https://github.com/Denis-Stanev/DNN-vs-BDT )}. \\
The given dataset is \texttt{MiniBooNE\_PID.txt}, a set of $130\,064$ reconstructed samples with $50$ features (most of statistical nature). From a preliminary analysis, we find that most feature's minimum values are extremely distant (up to two orders of magnitude) from the others, resulting in big deviations from the real mean values. Most of said values are a minima at $(-999.0)$, occurring a non-negligeable number of times in both signal and background. This lead us to think of them as a reconstruction error, or detector default value, to remove. Nevertheless we will work on both datasets. Other instances of radical points occur at feature number $20$: three other abnormal minima, and a maximum point five orders of magnitude above other point's distribution. They will also be dealt with, as shown in the plots.

\newcommand{\scales}{0.4}
\begin{figure}[p]
    \centering
    \subfloat[][]{\hspace*{-2cm}\includegraphics[scale=\scales]{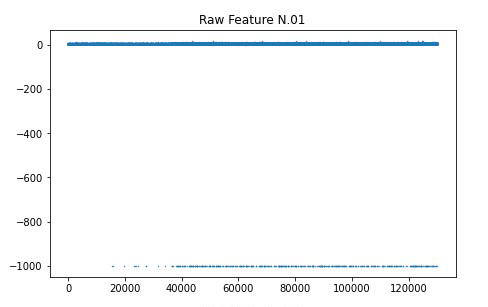}}
    \subfloat[][]{\includegraphics[scale=\scales]{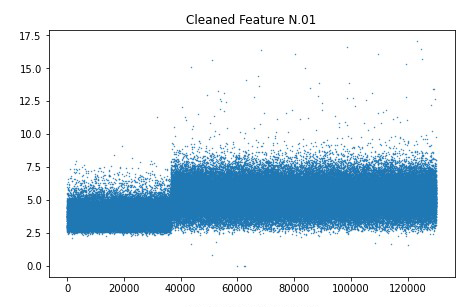}}
    \subfloat[][]{\includegraphics[scale=\scales]{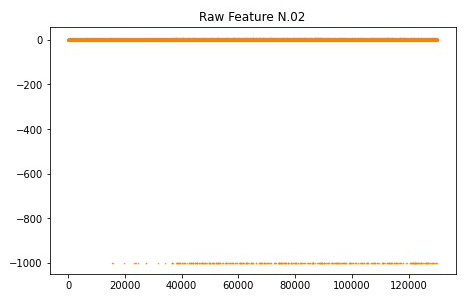}}
    \subfloat[][]{\includegraphics[scale=\scales]{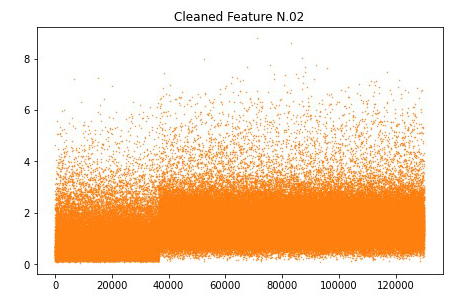}}
    \\
    \subfloat[][]{\hspace*{-2cm}\includegraphics[scale=\scales]{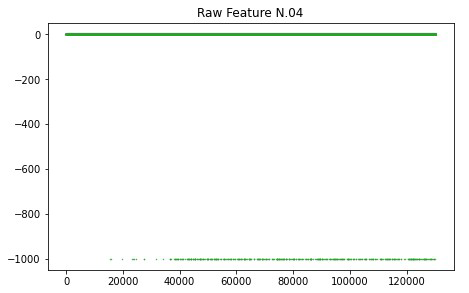}}
    \subfloat[][]{\includegraphics[scale=\scales]{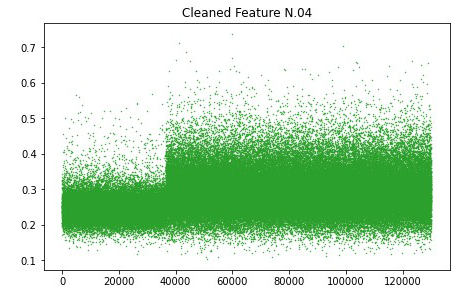}}
    \subfloat[][]{\includegraphics[scale=\scales]{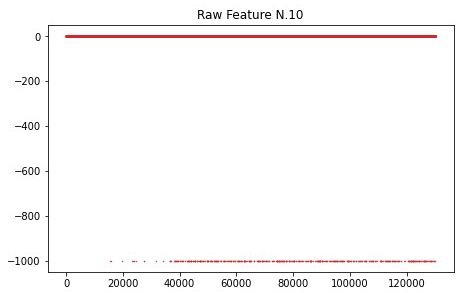}}
    \subfloat[][]{\includegraphics[scale=\scales]{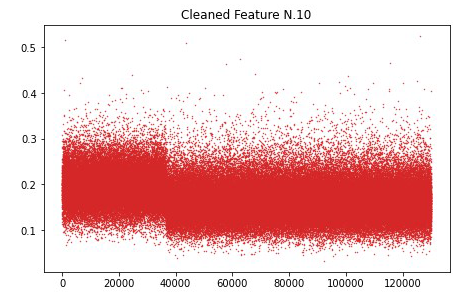}}
    \\
    \subfloat[][]{\hspace*{-2cm}\includegraphics[scale=\scales]{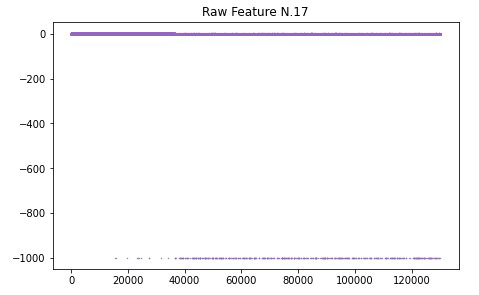}}
    \subfloat[][]{\includegraphics[scale=\scales]{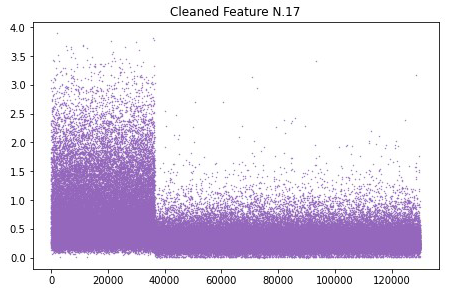}}
    \subfloat[][]{\includegraphics[scale=\scales]{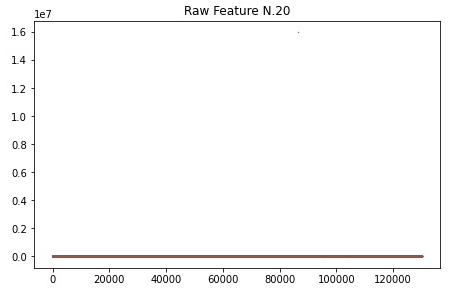}}
    \subfloat[][]{\includegraphics[scale=\scales]{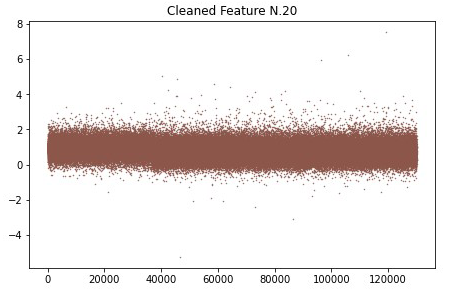}}
    \\
    \subfloat[][]{\hspace*{-2.2cm}\includegraphics[scale=\scales]{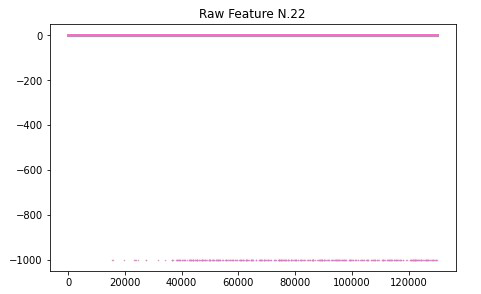}}
    \subfloat[][]{\includegraphics[scale=\scales]{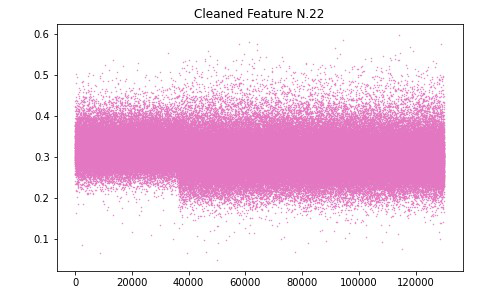}}
    \subfloat[][]{\includegraphics[scale=\scales]{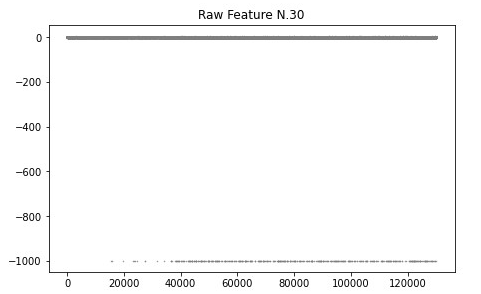}}
    \subfloat[][]{\includegraphics[scale=\scales]{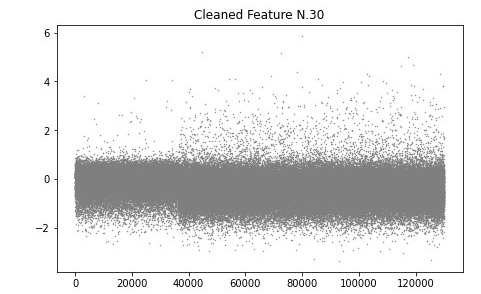}}
    \\
    \subfloat[][]{\hspace*{-2.2cm}\includegraphics[scale=\scales]{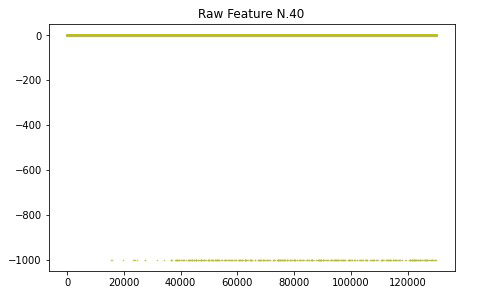}}
    \subfloat[][]{\includegraphics[scale=\scales]{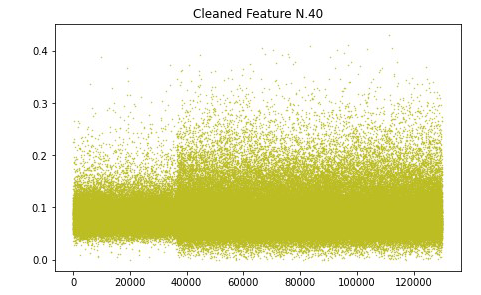}}
    \subfloat[][]{\includegraphics[scale=\scales]{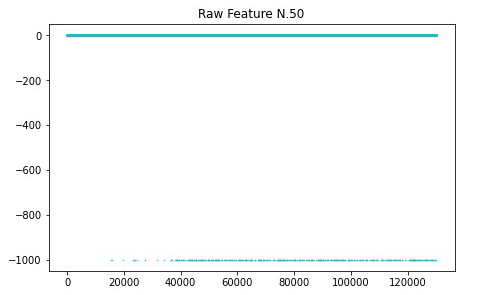}}
    \subfloat[][]{\includegraphics[scale=\scales]{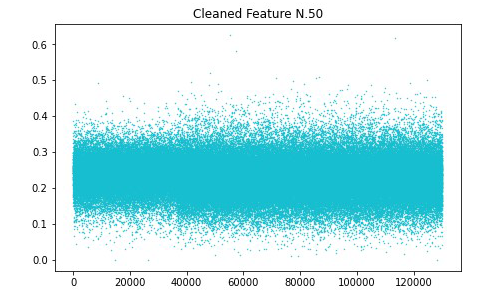}}
    \caption{Some of the features plots without and with data correction. As we can see the minimum values are far enough to flatten the scatter. It can especially be seen with feature 20's maximum ((k) and (l) brown graphs).}
    \label{fig:features}
\end{figure}

\begin{table}[ht]
    \centering
    \begin{tabular}{lrrrrrrr}
        \toprule
        $F_i$ & min($F_i$) & $2^{nd} min(F_i)$ & max($F_i$) & $mean(F_i)_{raw}$ & $mean(F_i)_{clean}$ & std$(F_i)_{raw}$ & std$(F_i)_{clean}$ \\
        \midrule
        $1$ & $-999.0$ & $0.0$ & $17.057$ & $1.252$ & $4.864$ & $60.12$ & $1.159$ \\
        $2$ & $-999.0$ & $0.06$ & $8.803$ & $-2.108$ & $1.492$ & $59.913$ & $0.859$ \\
        $3$ & $-999.0$ & $0.012$ & $4747.67$ & $123.795$ & $127.85$ & $196.01$ & $184.031$ \\
        $4$ & $-999.0$ & $0.105$ & $0.737$ & $-3.313$ & $0.283$ & $59.834$ & $0.059$ \\
        $5$ & $-999.0$ & $-0.156$ & $0.179$ & $-3.585$ & $0.009$ & $59.818$ & $0.012$ \\
        $\vdots$ & $\vdots$ & $\vdots$ & $\vdots$ & $\vdots$ & $\vdots$ & $\vdots$ & $\vdots$ \\
        $20$ & $-17257$ & $-12790.0$ & $119.90$ & $7.54$ & $1.60\per10^7$ & $44367.69$ & $0.467$ \\
        $\vdots$ & $\vdots$ & $\vdots$ & $\vdots$ & $\vdots$ & $\vdots$ & $\vdots$ & $\vdots$ \\
        $46$ & $-999.0$ & $0.0$ & $0.606$ & $-3.416$ & $0.179$ & $59.828$ & $0.08$ \\
        $47$ & $-999.0$ & $-32.11$ & $1.933$ & $-3.753$ & $-0.159$ & $59.816$ & $0.976$ \\
        $48$ & $-999.0$ & $-13.492$ & $21.768$ & $-0.745$ & $2.86$ & $60.019$ & $1.914$ \\
        $49$ & $-999.0$ & $-15.998$ & $25.423$ & $-1.987$ & $1.613$ & $59.946$ & $1.957$ \\
        $50$ & $-999.0$ & $0.0$ & $0.625$ & $-3.365$ & $0.231$ & $59.831$ & $0.053$ \\
        \bottomrule
    \end{tabular}
    \caption{Extract of the corrupted data in \texttt{MiniBooNE\_PID.txt}. $F_i$ states for $i$-th feature, columns two and three are the first and second minimum value for such feature. Then mean value per feature (original and not), first max value and Standard Deviation (before and after cleaning). The $20th$ feature has a really unusual distribution, as we can see from these values and from the Figure \ref{fig:features}(k). Once cleaned it also shows very little separation power. 
    The complete table can be found in the Appendix.}
    \label{tab:inconsistent_data}
\end{table}

\par The chosen method for removing these inconsistencies is chosen to be the least influent one. Since the data is fuzzy, we choose to substitute every spurious point with the feature's average taken without them. This is done for both the minima and feature 20's maximum. The mean without said points is computed over the whole feature, as splitting the set by label would mean biasing its distribution.

\clearpage
\section{PCA}
\begin{figure}[!h]
    \centering
    \subfloat[]{\hspace*{-1cm}\includegraphics[scale = 0.3]{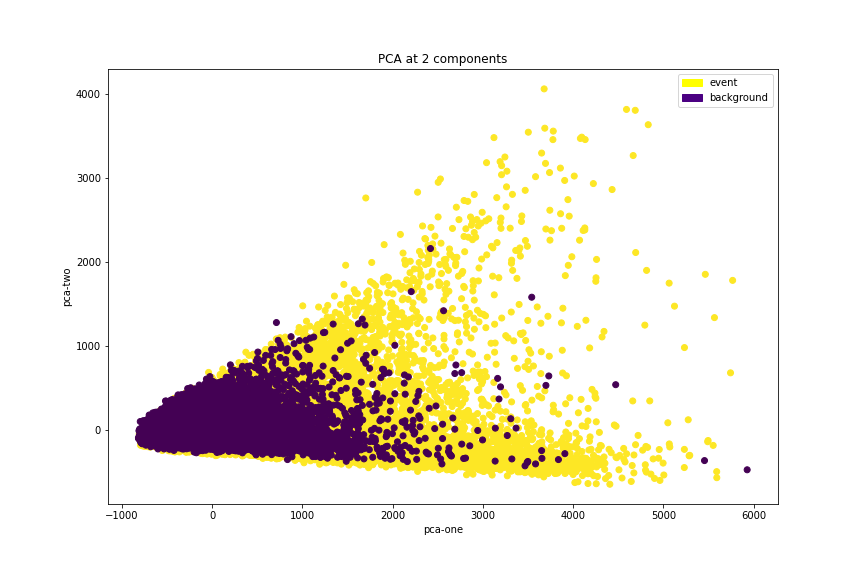}}
    \subfloat[]{\includegraphics[scale = 0.3]{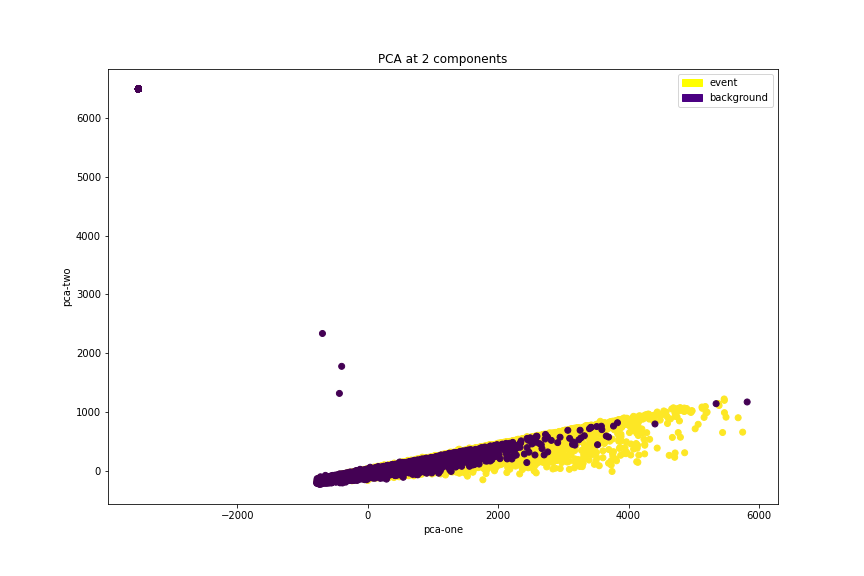}}
    \caption{Reduction of dimensions with a PCA, as we can see there is no significant distinction with two dimensions with either clean (a) and uncleaned (b) data (although we had to remove the point with value around $1e7$).}
    \label{fig:pca}
\end{figure}

We also tried to see the nature of the features reducing the dimensions in a PCA. We used $2$ dimensions but as we can see in Figure \ref{fig:pca}, the result is not satisfying. The algorithm doesn't separate the classes in a clear way, however we can see an improvement when the cleaned dataset is used. We'll then be using all the 50 features to define the respective classes.
\section{\textbf{BDT}: Boosted Decision Tree}
In order to improve the particle identification power we need the starting point of a BDT. We recreate the model used in the paper\cite{2005NIMPA.543..577R} following their instructions, so we use AdaBoost from scikit-learn \cite{scikit-learn} with a tree classifier with $\texttt{max\_leaf\_nodes}=45$, $1000$ estimators and a learning rate of $0.5$. The training takes a bit more than $1$ hour on Google Colab\link{https://colab.research.google.com}, and the final accuracy is around $0.945$.
\begin{figure}[h!]
    \subfloat[]{\hspace*{-0.5cm}\includegraphics[scale=0.34]{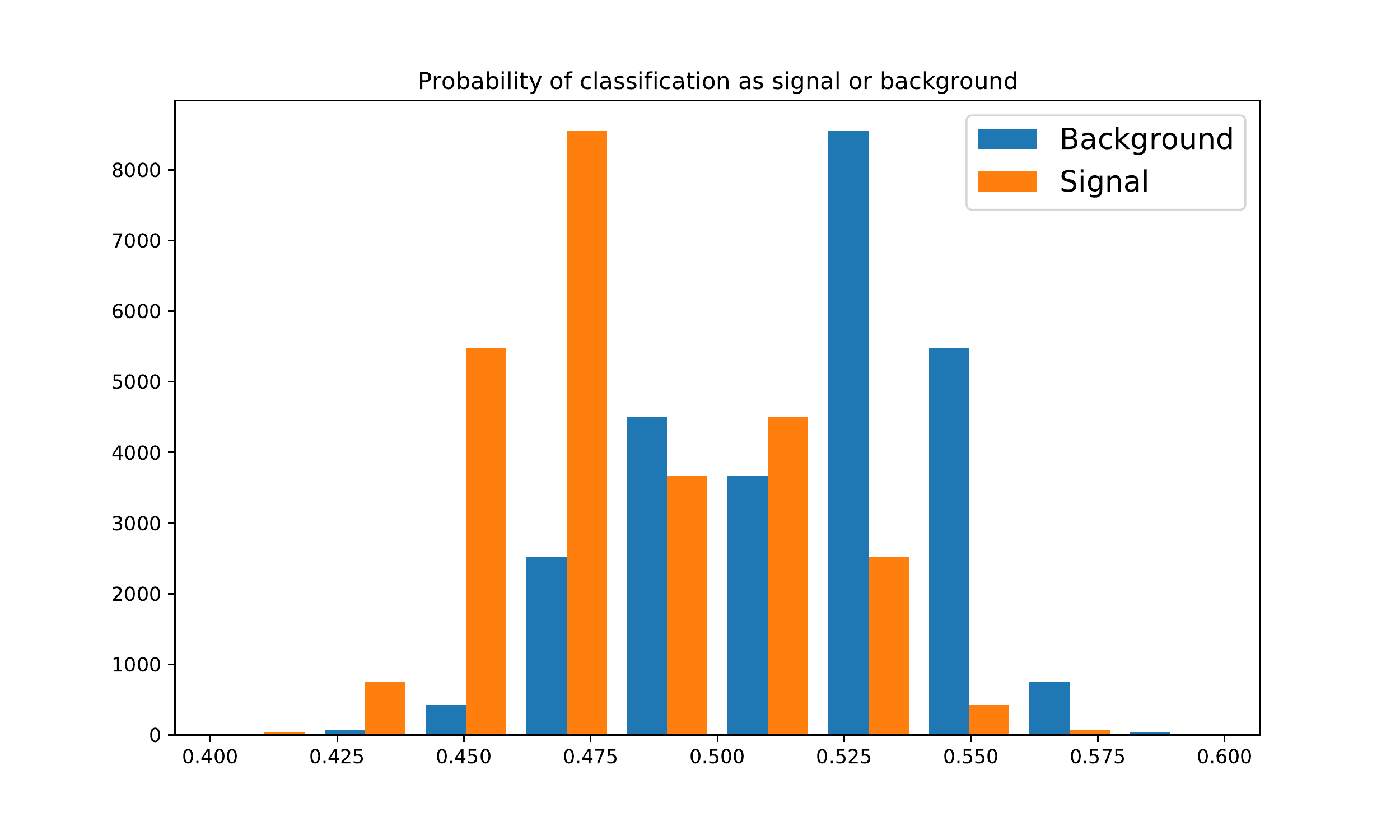}}
    \subfloat[]{\includegraphics[scale=0.34]{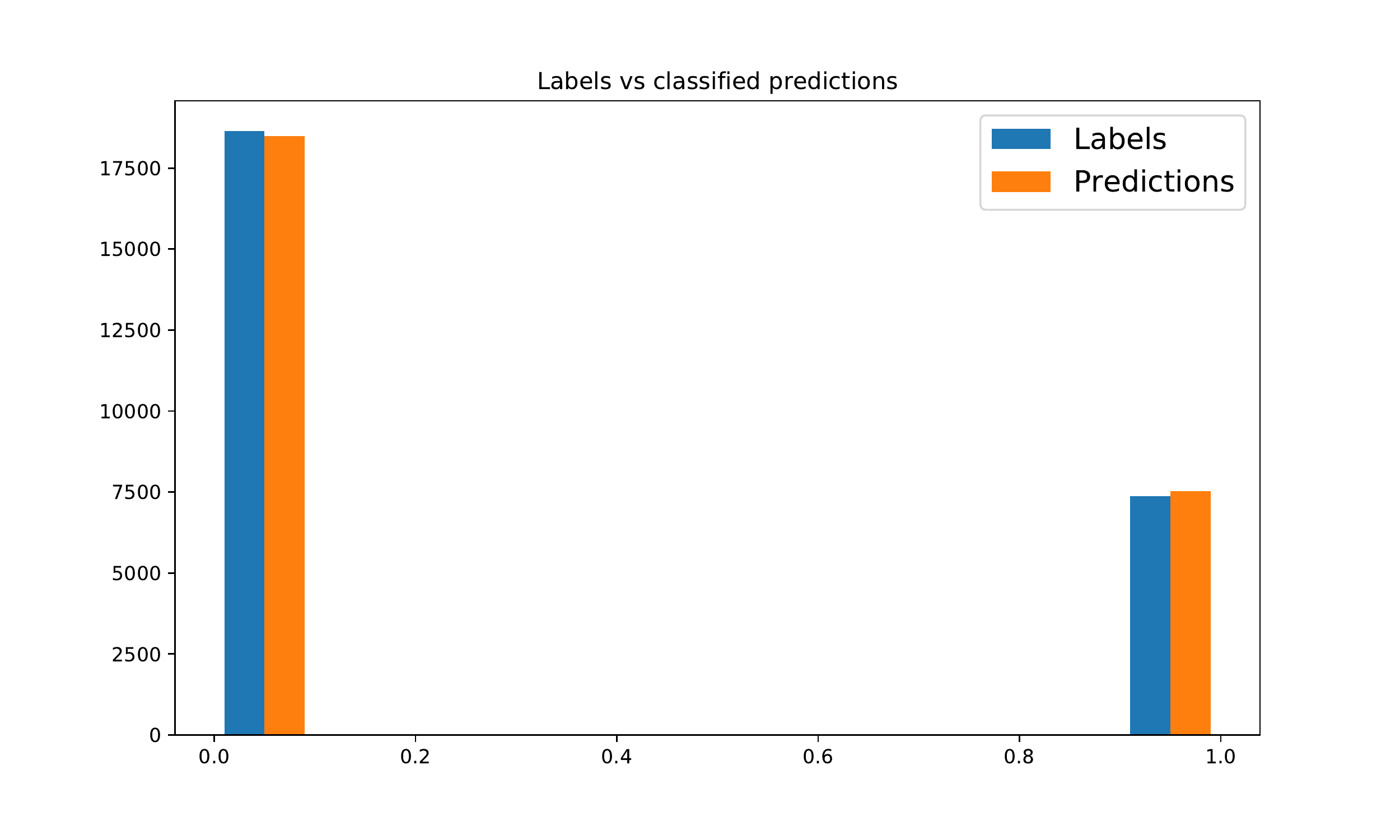}}
    \caption{Results from AdaBoost Decision Tree.}
    \label{fig:ada}
\end{figure}

\clearpage
\section{\textbf{DNN}: Deep Neural Network}
The main choice of network is between a deep \href{https://www.youtube.com/watch?v=ZiTVSjlTcvc}{MLP} (Multilayer perceptron)\cite{Aitkin2003} and a CNN (Convolutional neural network)\cite{LeCun1999}. We decided not to use any convolutional layers because the various features describe parameters that have different nature and very little in common, and as such, there is no translational invariance, which means there would have been no point in using convolutions. 
\par After many iterations, the selected DNN structure is the following: An input layer of size $50$, a fully connected hidden layer with $256$ nodes, two fully connected hidden layers with $128$ nodes each, one fully connected hidden layer with $64$ nodes, and the output, which has only a single node. The chosen activation functions for the hidden layers is ELU, and the sigmoid function for the output layer, since our outputs will be split between $0$ and $1$. The choice of loss function is trivial, and it is set to be Binary Crossentropy, minimized via Adam (its learning rate will be later discussed). Further testing has shown the need for dropout functions, which needed to be finely tuned through iterations. The final choice is a dropout with $0.2$ rate for the first $3$ hidden layers. We also normalize the data to a $0$ centered gaussian with variance $1$ before feeding it to the network, using the cleaned data described in section \ref{DataAnalysis}.
\begin{figure}[ht]
    \centering
    \includegraphics[scale=0.3]{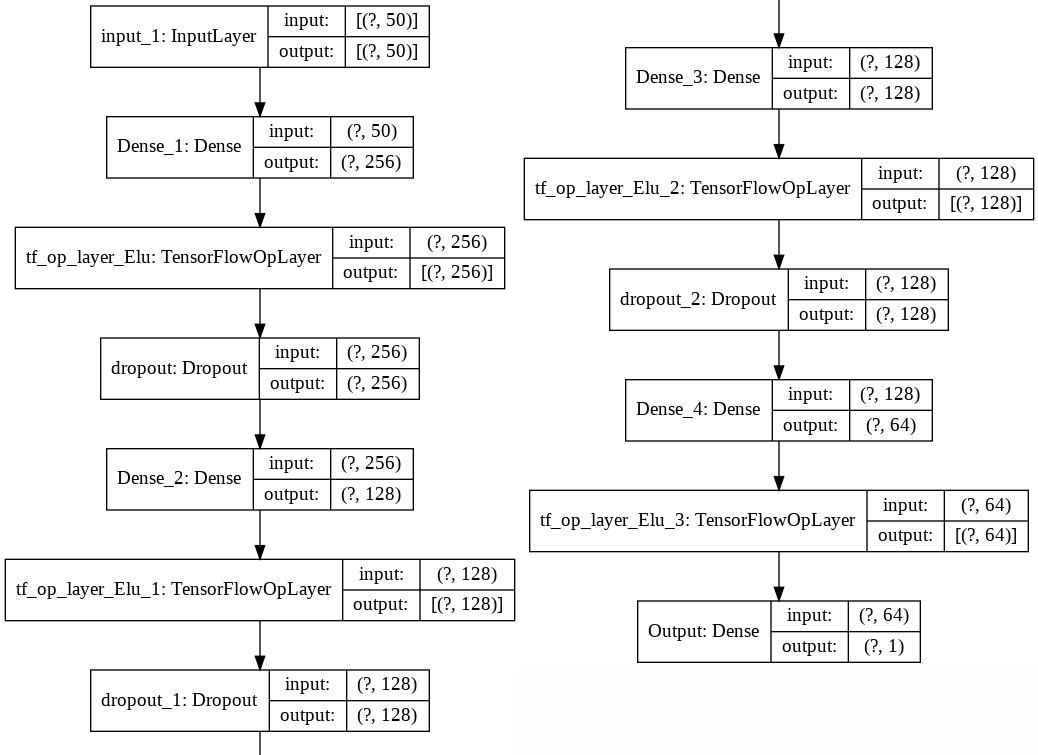}
    \label{fig:model}
    \caption{Scheme of the designed model. The model was created using Tensorflow\cite{tensorflow}. The TF operative layer is where we use the ELU activation function for all $4$ blocks.}
\end{figure}

\par
\begin{figure}[h]
    \centering
    \includegraphics[scale=0.4]{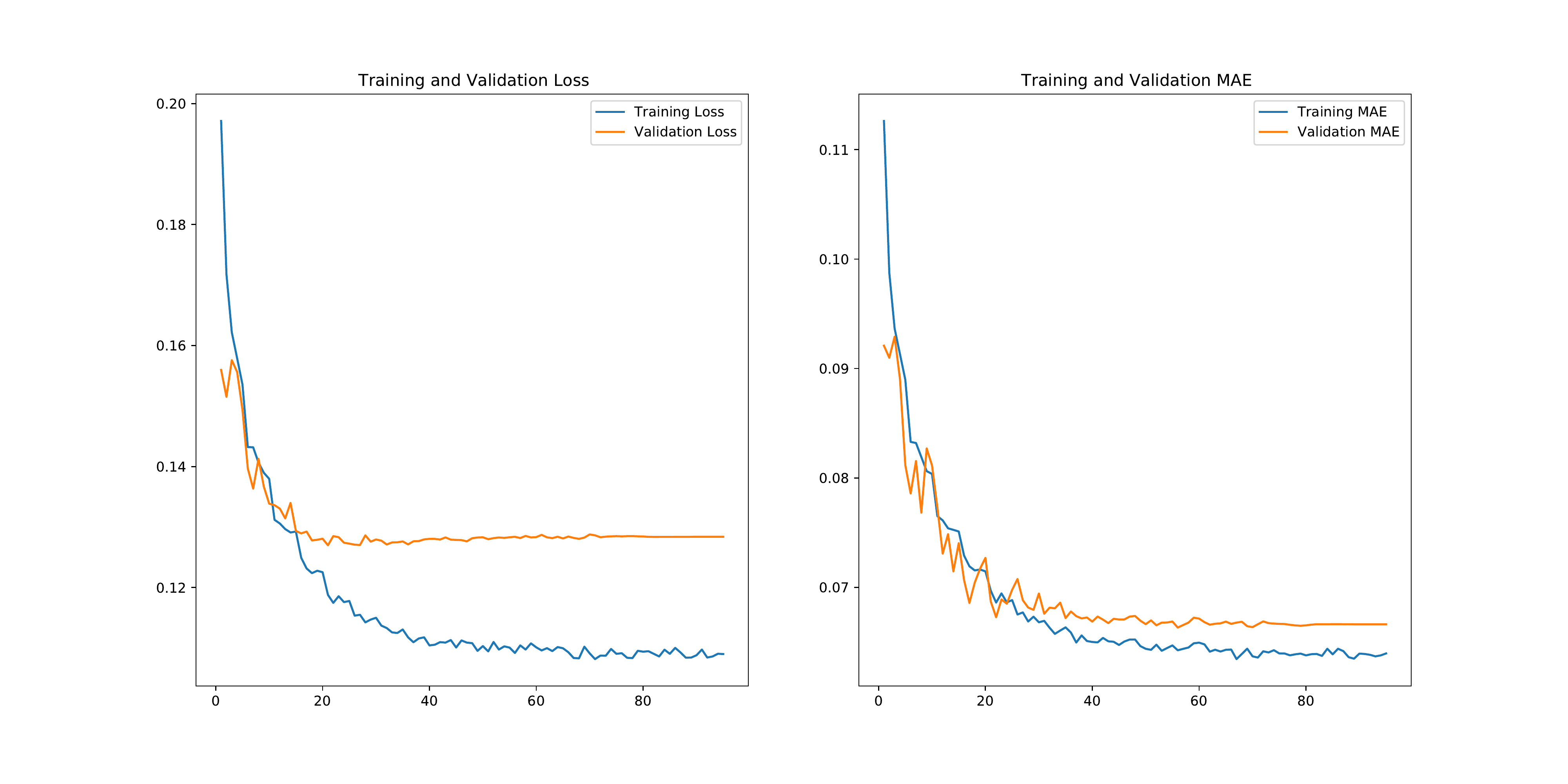}
    \caption{Loss and MAE performance during training of the model.}
    \label{fig:lossDNN}
\end{figure}

 As for the learning rate, we keep it fixed at first, and use the technique discussed in this paper\cite{BatchSize} in which batch size is procedurally increased, rather than decreasing the learning rate. The learning convergence is the same, and it led to the same accuracy, but in a shorter training time, as the training speed increases with bigger batch sizes (this increase in speed strongly depends on how good the GPUs are). We train with a learning rate of $0.001$ for $45$ epochs, starting at a batch size of $32$ and doubling it every $5$ epochs, ending at a batch size of $8192$. We then train for an additional $50$ epochs at batch size $8192$ with decaying learning rate to fine tune the weights around the found minimum.
In figure \ref{fig:lossDNN} we can see the loss and mean absolute error on the training and validation sample.

\par At the end of training we predict our test set labels, and since such predictions are continuous, we split them in between: greater than $0.5$ goes to $1$, and the rest to $0$. We finally get an accuracy of around $0.952$ (the value of the accuracy fluctuates between $0.950$ and $0.954$ depending on the run and the split dataset fed into the net, but on average it is $0.952$), and the training time is between $100$ and $300$ seconds, depending on which GPU we get on Colab (usually around 150 seconds).

\par From figure \ref{fig:lossDNN} it is clear that the learning method was a success: both loss and validation loss converge somewhat smoothly, and do not differ too much in their asymptotical values, and there seems to be no sign of overfitting (also thanks to dropout). 
\par It should be noted that it doesn't make sense to split between labels at values different from $0.5$; even though in some runs the accuracy can be greater at different values, choosing one would mean training on the test set, which is an ill practice for problem generalisation.
\par
\begin{figure}[h]
    \centering
    \includegraphics[scale=0.3]{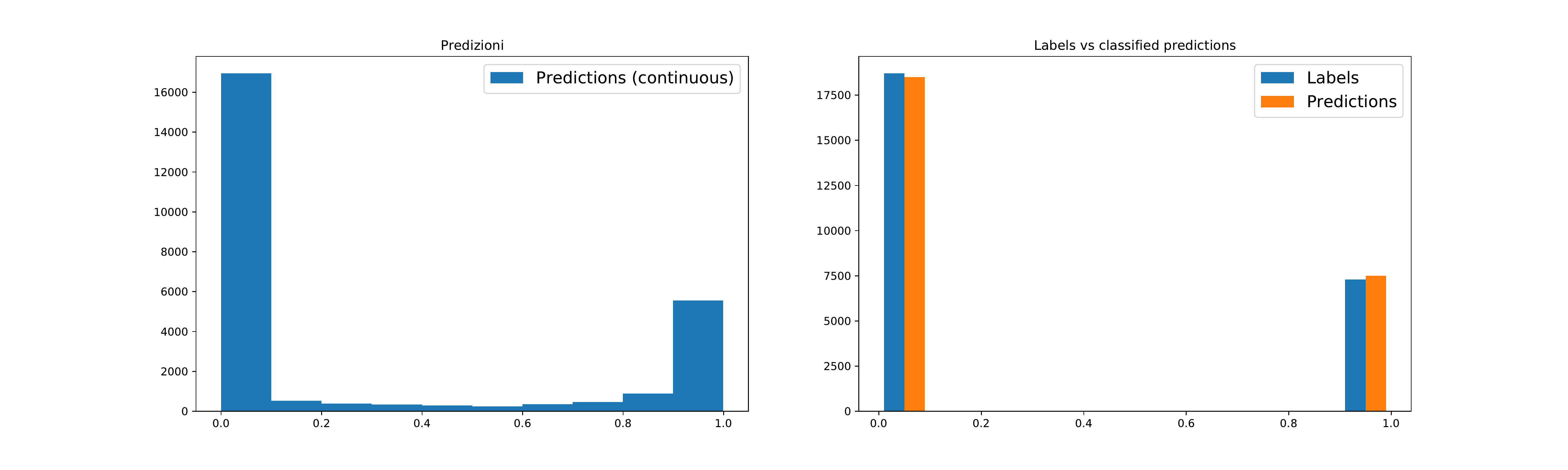}
    \caption{Histograms of prediction probabilities for our NN. From Figure \ref{fig:ada} we can see that our Neural Network maximised the probability at $1$ or $0$, while the BDT concentrates the probabilities near the middle.}
    \label{fig:predHist}
\end{figure}

 In figure \ref{fig:predHist} we can see how the prediction probabilities are distributed.

\newpage
\section{Results comparisons}
 As we can see, with our neural network we achieved a slight improvement in accuracy, going from around $ 0.945 $ to about $ 0.952 $, and a big improvement in training time, as the network takes less than one tenth of the time it takes to train the AdaBoost algorithm.
 \par In figures \ref{fig:ada} and \ref{fig:predHist} we can also see how the probability distributions differ from the two methods, with the Adaboost trees being grouped near the $0.5$ value, while the DNN puts most values at either $0$ or $1$ with very few values classified in between the two. This shows that the neural network has a higher confidence in its predictions, because the loss function maximises the output value to be as close as possible to the real value, while the Boosted Decision Tree only cares about classifying the point correctly, and not maximising its certainty about said classification.

\section{Conclusions}
In the last decade there have been many improvements in deep learning techniques, and using some of these we were able to create a neural network that performed better and trained faster than the one in the paper\cite{2005NIMPA.543..577R} from 2004. Nevertheless we weren't able to achieve a noticeably higher accuracy. This led us to think that the dataset has an accuracy limit and we can't go further because, as stated in the original paper, these features are the result of the reconstruction package R-fitter which, at the time, was still undergoing modifications to improve its feature generation algorithms. We think that one way to obtain better results would be to use data directly from the particle detectors instead of the reconstructed ones, as some important information might be lost in the reconstruction. Today's machine learning techniques and better computational power would allow for the creation of neural networks that can handle successfully more complex problems with many more input features than what was available at the time of the original paper.

\newpage

\bibliographystyle{unsrt}
\bibliography{main}

\begin{thebibliography}{1}

\bibitem{2005NIMPA.543..577R}
Byron~P. {Roe}, Hai-Jun {Yang}, Ji~{Zhu}, Yong {Liu}, Ion {Stancu}, and Gordon
  {McGregor}.
\newblock {Boosted decision trees as an alternative to artificial neural
  networks for particle identification}.
\newblock {\em Nuclear Instruments and Methods in Physics Research A},
  543(2-3):577--584, May 2005.

\bibitem{Iiyama_2021}
Yutaro Iiyama, Gianluca Cerminara, Abhijay Gupta, Jan Kieseler, Vladimir
  Loncar, Maurizio Pierini, Shah~Rukh Qasim, Marcel Rieger, Sioni Summers,
  Gerrit Van~Onsem, and et~al.
\newblock Distance-weighted graph neural networks on fpgas for real-time
  particle reconstruction in high energy physics.
\newblock {\em Frontiers in Big Data}, 3, Jan 2021.

\bibitem{Kieseler_2020}
Jan Kieseler.
\newblock Object condensation: one-stage grid-free multi-object reconstruction
  in physics detectors, graph, and image data.
\newblock {\em The European Physical Journal C}, 80(9), Sep 2020.

\bibitem{scikit-learn}
F.~Pedregosa, G.~Varoquaux, A.~Gramfort, V.~Michel, B.~Thirion, O.~Grisel,
  M.~Blondel, P.~Prettenhofer, R.~Weiss, V.~Dubourg, J.~Vanderplas, A.~Passos,
  D.~Cournapeau, M.~Brucher, M.~Perrot, and E.~Duchesnay.
\newblock Scikit-learn: Machine learning in {P}ython.
\newblock {\em Journal of Machine Learning Research}, 12:2825--2830, 2011.

\bibitem{Aitkin2003}
Murray Aitkin and Rob Foxall.
\newblock Statistical modelling of artificial neural networks using the
  multi-layer perceptron.
\newblock {\em Statistics and Computing}, 13(3):227--239, Aug 2003.

\bibitem{LeCun1999}
Yann LeCun, Patrick Haffner, L{\'e}on Bottou, and Yoshua Bengio.
\newblock {\em Object Recognition with Gradient-Based Learning}, pages
  319--345.
\newblock Springer Berlin Heidelberg, Berlin, Heidelberg, 1999.

\bibitem{tensorflow}
Mart\'{\i}n Abadi, Ashish Agarwal, Paul Barham, Eugene Brevdo, Zhifeng Chen,
  Craig Citro, Greg~S. Corrado, Andy Davis, Jeffrey Dean, Matthieu Devin,
  Sanjay Ghemawat, Ian Goodfellow, Andrew Harp, Geoffrey Irving, Michael Isard,
  Yangqing Jia, Rafal Jozefowicz, Lukasz Kaiser, Manjunath Kudlur, Josh
  Levenberg, Dandelion Man\'{e}, Rajat Monga, Sherry Moore, Derek Murray, Chris
  Olah, Mike Schuster, Jonathon Shlens, Benoit Steiner, Ilya Sutskever, Kunal
  Talwar, Paul Tucker, Vincent Vanhoucke, Vijay Vasudevan, Fernanda Vi\'{e}gas,
  Oriol Vinyals, Pete Warden, Martin Wattenberg, Martin Wicke, Yuan Yu, and
  Xiaoqiang Zheng.
\newblock {TensorFlow}: Large-scale machine learning on heterogeneous systems,
  2015.
\newblock Software available from tensorflow.org.

\bibitem{BatchSize}
Samuel~L. Smith, Pieter-Jan Kindermans, Chris Ying, and Quoc~V. Le.
\newblock {Don't Decay the Learning Rate, Increase the Batch Size}.
\newblock {\em ICLR 2018}, 2018.

\end{thebibliography}

\appendix\section{Data Table}
In the next page the complete\footnote{Partial table is here \ref{tab:inconsistent_data}} table with all the features.
\begin{table}[]
    \centering
    \begin{tabular}{lrrrrrrr}
        \toprule
        $F_i$ & min($F_i$) & $2^{nd} min(F_i)$ & max($F_i$) & $mean(F_i)_{raw}$ & $mean(F_i)_{clean}$ &      std$(F_i)_{raw}$ & std$(F_i)_{clean}$ \\

        \midrule
        $1$ & $-999.0$ & $0.0$ & $17.057$ & $1.252$ & $4.864$ & $60.12$ & $1.159$ \\
        $2$ & $-999.0$ & $0.06$ & $8.803$ & $-2.108$ & $1.492$ & $59.913$ & $0.859$ \\
        $3$ & $-999.0$ & $0.012$ & $4747.67$ & $123.795$ & $127.85$ & $196.01$ & $184.031$ \\
        $4$ & $-999.0$ & $0.105$ & $0.737$ & $-3.313$ & $0.283$ & $59.834$ & $0.059$ \\
        $5$ & $-999.0$ & $-0.156$ & $0.179$ & $-3.585$ & $0.009$ & $59.818$ & $0.012$ \\
        $6$ & $-999.0$ & $0.0$ & $0.704$ & $-3.434$ & $0.161$ & $59.827$ & $0.123$ \\
        $7$ & $-999.0$ & $0.0$ & $6.241$ & $-2.628$ & $0.97$ & $59.876$ & $0.342$ \\
        $8$ & $-999.0$ & $0.034$ & $0.99$ & $-2.774$ & $0.824$ & $59.867$ & $0.07$ \\
        $9$ & $-999.0$ & $2.375$ & $7.17$ & $-0.156$ & $3.451$ & $60.025$ & $0.258$ \\
        $10$ & $-999.0$ & $0.033$ & $0.525$ & $-3.421$ & $0.174$ & $59.828$ & $0.044$ \\
        $11$ & $-999.0$ & $0.0$ & $9.559$ & $0.71$ & $4.32$ & $60.077$ & $0.34$ \\
        $12$ & $-999.0$ & $-6.964$ & $537.262$ & $161.447$ & $165.638$ & $134.25$ & $114.717$ \\
        $13$ & $-999.0$ & $-1.0$ & $1.0$ & $-3.145$ & $0.451$ & $59.847$ & $0.522$ \\
        $14$ & $-999.0$ & $0.0$ & $1.0$ & $-3.285$ & $0.31$ & $59.836$ & $0.242$ \\
        $15$ & $-999.0$ & $0.959$ & $11.296$ & $-1.353$ & $2.25$ & $59.956$ & $0.701$ \\
        $16$ & $-999.0$ & $135.279$ & $6907.69$ & $933.035$ & $940.012$ & $652.824$ & $642.417$ \\
        $17$ & $-999.0$ & $0.003$ & $3.902$ & $-3.141$ & $0.455$ & $59.846$ & $0.416$ \\
        $18$ & $-999.0$ & $0.628$ & $1097.13$ & $23.311$ & $27.003$ & $68.403$ & $30.081$ \\
        $19$ & $-999.0$ & $0.057$ & $0.759$ & $-3.327$ & $0.268$ & $59.833$ & $0.052$ \\
        $20$ & $-17256.7$ & $-12790.0$ & $16000900.0$ & $119.908$ & $0.788$ & $44367.518$ & $0.467$ \\
        $21$ & $-999.0$ & $-23.588$ & $22.404$ & $-3.695$ & $-0.1$ & $59.841$ & $1.875$ \\
        $22$ & $-999.0$ & $0.049$ & $0.597$ & $-3.287$ & $0.309$ & $59.836$ & $0.048$ \\
        $23$ & $-999.0$ & $0.0$ & $1428.59$ & $80.653$ & $84.552$ & $106.955$ & $85.03$ \\
        $24$ & $-999.0$ & $1.164$ & $78.363$ & $0.238$ & $3.847$ & $60.075$ & $1.798$ \\
        $25$ & $-999.0$ & $0.0$ & $0.5$ & $-3.156$ & $0.441$ & $59.844$ & $0.058$ \\
        $26$ & $-999.0$ & $-39.63$ & $4.717$ & $-4.194$ & $-0.602$ & $59.818$ & $2.083$ \\
        $27$ & $-999.0$ & $0.0$ & $1024.21$ & $97.404$ & $101.363$ & $98.222$ & $72.845$ \\
        $28$ & $-999.0$ & $0.756$ & $3.647$ & $-2.034$ & $1.566$ & $59.912$ & $0.259$ \\
        $29$ & $-999.0$ & $0.0$ & $0.653$ & $-3.422$ & $0.174$ & $59.828$ & $0.055$ \\
        $30$ & $-999.0$ & $-3.352$ & $5.87$ & $-3.899$ & $-0.306$ & $59.802$ & $0.595$ \\
        $31$ & $-999.0$ & $2.394$ & $23.933$ & $3.952$ & $7.574$ & $60.315$ & $2.305$ \\
        $32$ & $-999.0$ & $-1.555$ & $4.127$ & $-2.583$ & $1.015$ & $59.879$ & $0.387$ \\
        $33$ & $-999.0$ & $0.003$ & $7.994$ & $-2.382$ & $1.217$ & $59.895$ & $0.756$ \\
        $34$ & $-999.0$ & $6.336$ & $499.999$ & $361.939$ & $366.853$ & $126.437$ & $96.426$ \\
        $35$ & $-999.0$ & $0.0$ & $0.403$ & $-3.522$ & $0.073$ & $59.822$ & $0.043$ \\
        $36$ & $-999.0$ & $0.043$ & $0.897$ & $-3.3$ & $0.296$ & $59.835$ & $0.069$ \\
        $37$ & $-999.0$ & $-1.656$ & $8.192$ & $-3.205$ & $0.391$ & $59.847$ & $0.853$ \\
        $38$ & $-999.0$ & $-13.166$ & $0.67$ & $-6.365$ & $-2.78$ & $59.659$ & $0.996$ \\
        $39$ & $-999.0$ & $0.144$ & $6.059$ & $-2.318$ & $1.281$ & $59.895$ & $0.288$ \\
        $40$ & $-999.0$ & $0.0$ & $0.43$ & $-3.503$ & $0.091$ & $59.823$ & $0.041$ \\
        $41$ & $-999.0$ & $33.895$ & $331.925$ & $141.18$ & $145.298$ & $75.499$ & $31.709$ \\
        $42$ & $-999.0$ & $-387.617$ & $161.298$ & $-25.226$ & $-21.709$ & $67.576$ & $33.797$ \\
        $43$ & $-999.0$ & $-1.686$ & $43.651$ & $-2.864$ & $0.733$ & $59.874$ & $1.222$ \\
        $44$ & $-999.0$ & $0.249$ & $85.823$ & $2.393$ & $6.009$ & $60.349$ & $4.554$ \\
        $45$ & $-999.0$ & $0.0$ & $0.447$ & $-3.586$ & $0.009$ & $59.818$ & $0.029$ \\
        $46$ & $-999.0$ & $0.0$ & $0.606$ & $-3.416$ & $0.179$ & $59.828$ & $0.08$ \\
        $47$ & $-999.0$ & $-32.11$ & $1.933$ & $-3.753$ & $-0.159$ & $59.816$ & $0.976$ \\
        $48$ & $-999.0$ & $-13.492$ & $21.768$ & $-0.745$ & $2.86$ & $60.019$ & $1.914$ \\
        $49$ & $-999.0$ & $-15.998$ & $25.423$ & $-1.987$ & $1.613$ & $59.946$ & $1.957$ \\
        $50$ & $-999.0$ & $0.0$ & $0.625$ & $-3.365$ & $0.231$ & $59.831$ & $0.053$ \\
        \bottomrule
    \end{tabular}
    \label{tab:complete_data}
\end{table}

\end{document}